\documentclass[conference]{IEEEtran}
\usepackage{cite}
\usepackage{amsmath,amssymb,amsfonts}
\usepackage{algorithmic}
\usepackage{svg}
\usepackage{url}
\usepackage{graphicx}
\graphicspath{{./images/}}
\usepackage{textcomp}
\usepackage{xcolor}
\def\BibTeX{{\rm B\kern-.05em{\sc i\kern-.025em b}\kern-.08em
    T\kern-.1667em\lower.7ex\hbox{E}\kern-.125emX}}
\usepackage[multiple]{footmisc}
\usepackage{booktabs}

\usepackage{etoolbox}
\makeatletter
\patchcmd{\@makecaption}
{\scshape}
{}
{}
{}
\makeatother

\begin{document}

IEEE Copyright Notice
Copyright (c) 2019 IEEE
Personal use of this material is permitted.  Permission from IEEE must be obtained for all other uses, in any current or future media, including reprinting/republishing this material for advertising or promotional purposes, creating new collective works, for resale or redistribution to servers or lists, or reuse of any copyrighted component of this work in other works.\\

\noindent To be published in:\\

\noindent Proceedings of the 2019 International Conference on Communications
(ICC’19: 20-24 May 2019, Shanghai, China)\\
DOI: to be published

\newpage

\topmargin=-0.5in

\title{\emph{affinity}: A System for Latent User Similarity Comparison on Texting Data 
}

\author{\IEEEauthorblockN{Tobias Eichinger, Felix Beierle}
\IEEEauthorblockA{\textit{Service-centric Networking} \\
\textit{Telekom Innovation Laboratories}\\
\textit{Technische Universit\"at Berlin}\\
Berlin, Germany \\
\{tobias.eichinger,beierle\}@tu-berlin.de}
\and
\IEEEauthorblockN{Sumsam U.\ Khan, Robin Middelanis,\\ Veeraraghavan Sekar, and Sam Tabibzadeh}
\textit{Technische Universit\"at Berlin}\\
Berlin, Germany \\
}

\maketitle

\begin{abstract}
In the field of social networking services, finding similar users based on profile data is common practice. Smartphones harbor sensor and personal context data that can be used for user profiling. Yet, one vast source of personal data, that is text messaging data, has hardly been studied for user profiling. We see three reasons for this: First, private text messaging data is not shared due to their intimate character. Second, the definition of an appropriate privacy-preserving similarity measure is non-trivial. Third, assessing the quality of a similarity measure on text messaging data representing a potentially infinite  set of topics is non-trivial. In order to overcome these obstacles we propose \emph{affinity}, a system that assesses the similarity between text messaging histories of users reliably and efficiently in a privacy-preserving manner. Private texting data stays on user devices and data for comparison is compared in a latent format that neither allows to reconstruct the comparison words nor any original private plain text. We evaluate our approach by calculating similarities between Twitter histories of 60 US senators. The resulting similarity network reaches an average 85.0\% accuracy on a political party classification task.
\end{abstract}


\section{Introduction}

Existing Social Networking Services (SNSs) are commonly used to connect people one already knows. They often suggest new contacts based on
user profile data or the social graph.
Especially with the proliferation of smartphones and the popularity
of location-based services, there is also a growing interest in more dynamic scenarios.
Combining the ideas of ubiquitous computing and social networking, it is possible to connect to similar and close by people who not necessarily are close in the social graph \cite{Beierle2018TrustCom,Talk2MeFrameworkDevicetoDevice2018}. In the field of psychology, the principle of \emph{homophily} states that individuals tend to bond if they are similar. This tendency also holds true within social networks services employing digital communication \cite{Beierle2017AIMS}.

Nowadays, a large portion of digital communication happens on smartphones which harbor highly personalized sensor and personal context data such as visited locations, music listened to, or app usage statistics \cite{Beierle2018MobiSPC}. Profiling users upon this data allows to estimate user similarities
\cite{Beierle2015HotPOST,Beierle2017AIMS}. A very commonly used -- yet virtually never mined -- resource on smartphones for user profiling  is text messaging\footnote{\url{http://www.pewinternet.org/2015/04/01/us-smartphone-use-in-2015/}} or simply \emph{texting}. When texting, users will cover topics of their interest, include personal information on their background and use individual vocabulary such as slang and emoticons. In particular, there is scientific evidence that personality is projected and perceived through digital communication \cite{Gill2003}. Hence, people with similar personality traits will tend to write similarly, which makes it eligible as a base for similarity estimation. However, a definition of similarity is not straightforward since most profiling approaches involve a full view on the data to derive similarity. Even if private texting data is promising as a meaningful basis for user profiling and similarity estimation, to the best of our knowledge, it essentially is uncharted terrain. 

We see three main problems: First, private text messaging data is not shared due to its intimate character. Second, the definition of an appropriate privacy-preserving similarity measure on texting data is non-trivial. Third, assessing the quality of a similarity measure on text messaging data representing a potentially infinite  set of topics is non-trivial. In particular, it is unclear which similarity benchmark to use. Should we evaluate on single benchmarks such as personality traits or user demographics such as gender, age, and political affiliation, or all together? In order to tackle these three problems we propose \emph{affinity}, a system for comparing user similarity on latent, that is in particular incomprehensible for humans, word vectors derived from a word embedding trained on texting data. We estimate similarity with the Word Mover's Distance (WMD) \cite{Kusner2015}, which allows to measure the distance between two sets of word vectors independently of any benchmark. Even though \emph{affinity} employs a central server harboring minimal amounts of global knowledge between users, similarity comparisons are performed decentrally, that is the central server neither knows the basis for comparison, nor the similarities between users, nor who compared with whom. In order to evaluate \emph{affinity}, we propose an evaluation scheme in the form of a political party classification task in order to measure \emph{affinity}'s performance. We use Twitter data of US politicians for which \emph{affinity} derives pairwise similarities. Finally, the pairwise similarities are translated into a similarity network which can be interpreted as knowledge-based nearest neighbor machine learning classifier.
Our main contributions are:
\begin{itemize}
	\item Design and implementation of \emph{affinity}, a system which leverages private texting data for real-time user similarity comparison, while preserving privacy.
	\item Evaluation of \emph{affinity}'s performance on Twitter data of 60 US politicians. \emph{affinity} reaches an average accuracy of 85.0\% on a political party classification task.
\end{itemize}

\section{Related Work}
\label{sec:relalted-work}
We see two main areas of related work converge in the field of social networking via user profiling.\\ 
\textbf{Profiling on smartphone data:} Previous works have proposed paradigms for finding similar people in one's vicinity leveraging sensor and context data. One of the first works utilizing mobile phones for this purpose follows the idea of exchanging identifiers of existing Online Social Networks (OSN) via Bluetooth in order to 
view public profiles of nearby users
\cite{whozthat-2008}. Other works enable users to broadcast short profiles or manually entered topics of interest over a short distance \cite{mobiclique-2009,e-shadow-2014,Talk2MeFrameworkDevicetoDevice2018}. There are also automated approaches, which utilize personal context data in the form of user profiles for comparison
\cite{Beierle2015HotPOST,BeierleMobileSoft2018,Beierle2018TrustCom}.\\
\textbf{Profiling on natural language resources:} In the field of Natural Language Processing (NLP), extracting a user's writing style and selection of topics which reflect the author's demographic and psychological properties is referred to as \emph{author profiling}. Author profiling includes classifying an author's gender or demographic information and has been largely exercised on Twitter data. Similarities to our proposed approach involve the usage of term and document weights in combination with Part-of-Speech (POS) n-grams \cite{Reddy2017}, and word embedding features combined with logistic regression \cite{Akhtyamova2017} or with a multimodal model for gender classification \cite{Patra2018}. Other related work aims at constructing thematic user profiles by clustering word embedding features of topic words for full-text item recommendation \cite{Alekseev2017}. Predicting  topics of interest or personal opinions on tweets has been covered in \cite{Raghuram2016}. A very recent work proposes to simultaneously and dynamically embed users and words even for streams of tweets \cite{Liang2018}. 

In the present paper, we propose an approach based on NLP techniques in order to profile users on private texting data in a social networking scenario such that private data stays on the user device.

\section{Fundamentals}
\label{sec:fundamentals}

\subsection{From Documents to Users}
Text is sequential data, that is \emph{one-dimensional} and \emph{directed}. Given an unfinished sentence such as
\begin{quote}
	\emph{An apple is} ...
\end{quote}
we may guess the next word in the sentence. The sentence may be continued in a meaningful way with the words \emph{edible}, \emph{green}, or \emph{big}, yet less so with the word \emph{blue}. We have guessed possible subsequent words to the sentence upon the preceding 3 words (3-gram) in the sentence. Since some words are more likely to follow than others, we see that text production can be interpreted as a (conditional) random process depending on close words.

Furthermore, text documents consist of interdependent sentences. Producing sentences, analogously to producing a subsequent word, is not an arbitrary process. In the related fields of text classification, text categorization, and image retrieval, topic models assume that documents have an underlying conceptual \emph{topic} \cite{Deerwester1990,Landauer1997,Blei2003,Blei2012} that can be described by a set of characteristic words. For example, when writing on the citric acid cycle, the probability of producing the term \emph{adenosine triphosphate} will be more probable than for example when writing a fairy tale. Thus the \emph{topicality} of a text document influences the writing process. 

In the following we will view a user's own part of his/her texting history as a distinct document. Most users have a diverse set of people they text with on probably an even more diverse set of topics. Hence texting histories will not harbor a single topic yet rather be a mixed lot of topics. We will call this mixture a person's \emph{state of mind}. We construe a user's state of mind as weighted yet discriminative set of words. We then consider people similar if they have similar states of mind, and else dissimilar. Since texting histories in plain text certainly are of private character, we will use word representations which neither allow to reconstruct the comparison words nor any original private plain text.

\subsection{On Document Similarity}
Many search engines and information retrieval techniques revolve around the isolation of characteristic words which yield much information on the topic of a search query or a document. Virtually unlimited possibilities to rephrase any given sentence into a topically and logically equivalent sentence makes it hard to exclusively map words to topics. Many words will be shared between topics. For example the occurrence of the word \emph{adenosine triphosphate} does not imply that the topic of the text document topic necessarily involves the citric acid cycle. Vice versa its absence does not imply that the topic of the text document excludes the citric acid cycle. In order to tackle this problem of uncertainty, NLP techniques provide means to map single words or entire text documents into semantic space. In semantic space, a word or a text document is represented by a vector wherein commonly the cosine distance is used to measure distances. However, there are many different means of measuring similarity between documents. See for instance \cite{Gomaa2013} for a survey on traditional approaches. More recently, with the advent of powerful word embedding models, a new method of estimating document similarity upon latent word features has emerged, which we deem to be fit for texting histories.

\subsection{On Word Embeddings}
Word embeddings, also known as Vector Space Models (VSM) \cite{Turney2010}, map words to vectors in real space by evaluating distributional context information. The \emph{context} of a word are other words in its proximity such as for example the two words to the left and right. Originally being high dimensional and sparse by nature, recent low-dimensional and dense word embedding algorithms \cite{Bojanowski2017,Pennington2014,Salle2016a} have proven to be similar in performance \cite{Levy2014}. State-of-the-art word embeddings trained on very large text corpora capture many semantic and syntactic similarities between words \cite{Mikolov2013a}, yet there does not seem to be a \emph{perfect word embedding} being able to capture similarities in a best way \cite{Schnabel2015}. There are also indications that domain-specific embeddings can perform better on domain-specific tasks \cite{Joshi2015}, which would suggest that word embeddings trained on corpora derived from for instance SMS messages perform better on linguistic tasks on SMS messages.

\subsection{Word Mover's Distance}

The Word Mover's Distance (WMD) introduced by Kusner et al. \cite{Kusner2015} is an application of the Earth Mover's Distance\footnote{more generally: \emph{Wasserstein distance}} (EMD) to word embeddings. The EMD measures the cost of moving earth from $i$ points to $j$ points in real space and may thus be reformulated as a transportation problem. Among others, the EMD has been traditionally used to compare grey-scale and color histograms of images for image comparison \cite{Ruzon2001, Grauman2004} and image retrieval \cite{Rubner2000}. Yet Kusner et al. \cite{Kusner2015} apply the EMD to word embeddings in order to compare sentences such as (A): "\emph{\textbf{Obama speaks} to the \textbf{media} in \textbf{Illinois}}." and (B): "\emph{The \textbf{President greets} the \textbf{press} in \textbf{Chicago}}.". They do so by dropping \emph{stop words} (non-bold), which are very common and thus do not convey content information, and afterwards moving the vector representations of the remaining words in sentence (A) to those in sentence (B). Word embeddings allocate word vectors of semantically similar words close to each other, for example the word vectors of the words \emph{Obama} and \emph{President} are close to each other. Thus, when we move one vector to the other, the cost is low / the distance is short. Now, when the words of two sentences are pairwise semantically similar, their pairwise costs are low / distances are short. The sentences are then assumed to convey messages on similar topics\footnote{Note that in general we cannot expect that the messages convey the same information.}. Instead of sentences, we will apply the WMD to entire texting histories. Using an entire texting history's vocabulary is computationally intractable. The WMD scales super cubic on the number of points to move \cite{Huang2016} and even small texting histories easily exceed a thousand vocabulary words. In order to circumvent this problem, we pick a small set of \emph{relevant} words for each and every user as a condensate of their states of mind, and compare similarity on this limited set of words. Note that the WMD has to be equipped with a designated distance function. Throughout our experiments, we employ the cosine distance\footnote{
$\cos(x,y) = x\cdot y / \Vert x \Vert \Vert y \Vert$, where $\cdot$ denotes the dot product.}
 since it is naturally bounded and commonly used with word vectors. For formal information on the underlying transportation problem and efficient solvers we refer to \cite{Kusner2015} and the references therein.

\subsection{tf-idf Features}

Term frequency$-$inverse document frequency (tf$-$idf) features are commonly used as relevance measures for text retrieval \cite{Salton1988, Beel2017}. It follows the rationale that frequently used words \emph{within} a document are important, yet they are only relevant if they are simultaneously less frequent \emph{over all} documents. We define tf$-$idf features for a term $t$ and a document $d$ by
\begin{equation}\label{tfidf}
\begin{array}{lll}
\text{tf-idf}(t,d) &= \text{tf}(t,d) &\cdot \text{idf}(t,d) \\
&= \frac{f_{t,d}}{ \max_{x\in {\cal{D}}}  f_{x,d}  } &\cdot  \log \frac{\vert {\cal{D}}\vert }{ \vert \{d \in {\cal{D}} : t \in d \} \vert },
\end{array}
\end{equation}
where $f_{t,d}$ is the raw count of occurrences of the term $t$ in the document $d$ and ${\cal{D}}$ is a collection of documents. Note that there are alternative definitions of tf$-$idf features \cite{Manning2008}. We choose this definition in order to mitigate a bias toward long documents.

\section{Concept}\label{concept}
\begin{figure}[htbp]
	\includegraphics[width=\linewidth,trim={10cm 0 10cm 0},clip=true]{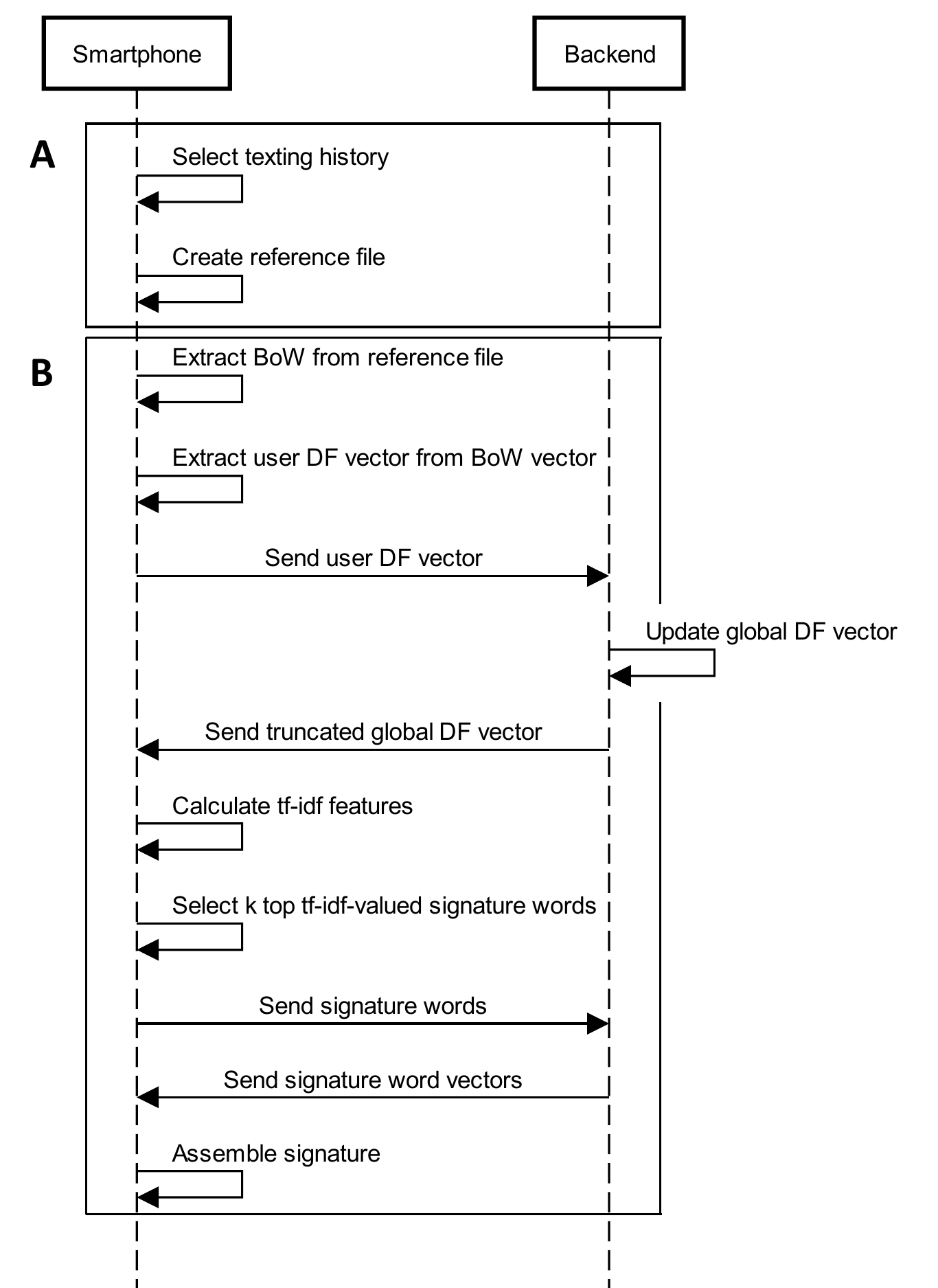}
	\caption{Message flow diagram of \emph{affinity}'s pre-calculation procedure before similarity comparison. The smartphone holds the plain texting history, and its signature. The backend holds the global DF vector and a reference word embedding model.}
	\label{fig:process_pipeline}
\end{figure}

We will describe \emph{affinity}'s similarity comparison along three main blocks, that is text selection, latent feature pre-calculation, and lastly similarity comparison upon request. For a seamless user experience, we designed an extended pre-calculation procedure (see Figure \ref{fig:process_pipeline}) paired with quick similarity comparisons.

\subsection{Text Selection}\label{text_insertion}
The user selects a texting history in plain text form upon which he/she wants to be compared with other users. All messages that are not self-written are dropped.  We call the resulting file \emph{reference file}. It is stored on the user's smartphone. 

\subsection{Latent Feature Pre-Calculation}\label{pre_calculation}
The reference file is transformed into a Bag-of-Words (BoW) vector, which is a dictionary with keys of unique vocabulary words and values of word occurrence counts.  The BoW vector is used to derive a user document frequency (DF) vector which shares the dictionary keys of the BoW vector, yet only holds ones in the values. The user DF vector is submitted to the backend, hence the backend only sees that the user has used a certain word yet not how often. The user DF vector updates the global DF vector that stores all words ever submitted and the number of their respective users. After updating the global DF vector, a truncated DF vector corresponding to the words submitted by the user, and the total number of users is returned to the smartphone. The user can now calculate tf$-$idf features for all of his/her used words according to Equation (\ref{tfidf}), which are then added to his/her BoW vector. The WMD calculates distances between two (word vectors, word weights) pairs, which are commonly referred to as \emph{signature}. Note that the WMD calculates distances on a word level and does only consider structural and distributional information such as word order that is condensed in the central word embedding. We see that the quality of similarity comparison hinges on the quality of the underlying word embedding.

In order to create user signatures, the top $k$ tf$-$idf valued words are submitted to the backend\footnote{The backend sees the user's $k$ words to compare similarity on. Submitting randomly selected additional words can sufficiently obfuscate this insight.}. The backend derives the submitted signature words' word vectors via some secret reference word embedding model and sends them to the user's smartphone\footnote{Techniques from differential privacy can be used to add jitter to the true word vectors to obfuscate knowledge on the hidden word embedding model.}. The signature word vectors (as a $k\times D$ matrix) and their corresponding normalized BoW vector (of length $k$) form the user's signature, where $D$ is the embedding dimension of the reference word embedding model.\\

\underline{Example: Signature Calculation}\\
Let $k=2$ be the signature size and $D = 3$ the dimension of the word embedding. Let $T$ be a texting history of some user $A$. In $T$, we drop all messages that have not been written by $A$ resulting in some reference file $R$. Let $V$ be the vocabulary of $R$ and $\text{\#}V$ its vocabulary size. We can now derive the user BoW vector $(\text{\#}v)_{v\in V}$, where $\text{\#}v$ is the number of occurrences of the vocabulary word $v$ in $R$. The user DF vector can then be formalized as $\text{DF}_{\text{user}}(A) = [1]_{v \in V}$. The user DF vector is sent to the backend, where it updates the global DF vector 
\begin{equation}
\text{DF}_{\text{global}} = \sum_{u\in U}\text{DF}_{\text{user}}(u),
\end{equation}
where $U$ is the set of registered users. We truncate the global DF vector along the vocabulary $V$, that is we only keep the entries associated to vocabulary words in $V$:
\begin{equation}
\text{DF}_{\text{trunc}}(A) = \text{DF}_{\text{global}|v \in V}.
\end{equation}
The truncated global DF vector together with the number of total users $\text{\#}U$ are sent back to the smartphone. We can now calculate the tf$-$idf values for all vocabulary words in $V$ according to Equation (\ref{tfidf}). We then pick the top $2$ tf$-$idf valued vocabulary words, for instance \emph{country} and \emph{nation}, and send them to the backend. The backend prompts two associated $3$-dimensional vectors $v_{\text{country}} = [-0.25 \text{ } 0.5 \text{ }0.75]$ and $v_{\text{nation}}=[-0.23 \text{ } 0.51 \text{ } 0.6]$. We find in the user BoW vector that $R$ contains $5$ and $6$ occurrences of the words \emph{country} and \emph{nation} respectively, which results in the normalized BoW weight vector $[5/11,6/11]$. User $A$'s signature is thus:

\[
\text{signature}(A) = \left[
\begin{array}{lll}
\begin{bmatrix}
[-0.25 & 0.50  & 0.75 ] \\
[-0.23 & 0.51 & 0.60  ] 
\end{bmatrix}, &

\begin{bmatrix}
5/11\\
6/11 
\end{bmatrix}
\end{array}
\right].
\]$\hfill \Box$

We point out that user signatures, which are the basis for similarity comparison, are snapshots. As both individual texting histories and the global DF vector evolve over time, signatures will have to be updated once in a while. Note that replacing the central word embedding only requires an update of signature vectors by all users, independent of the global DF vector.

\subsection{Similarity Comparison}\label{similarity_comparison}
Any two users who are ready for comparison can share their signatures locally without the backend knowing, and calculate their pairwise similarity via the WMD on their smartphones. Note that comparison partners cannot view their partner's comparison words since they have been converted into latent feature vectors. Thus, signatures can be shared at will as long as the reference word embedding model remains secret.

\section{Implementation}\label{implementation}
We use the technology bundle \emph{nginx}, \emph{gunicorn}, and \emph{Flask} to set up \emph{affinity} as a Web-application\footnote{Find \emph{affinity}'s code at: https://github.com/moxplayer/affinity.} written in Python on an Ubuntu backend. We will restrict ourselves to the \emph{fasttext} word embedding algorithm \cite{Bojanowski2017} as reference word embedding model for three reasons. First, \emph{fasttext} takes into account similarities and dissimilarities between morphologically related words (for example \emph{entangle}, \emph{entangled}, \emph{disentangled}). Second, it embeds \emph{out-of-vocabulary} (OOV) words, that is words that have not been seen during training time. Third, it performs language identification in order to -- for example -- filter out non-English conversations.

Note that in this prototype setup, in contrast to \emph{affinity}'s concept, all data processing takes place on the backend, whereas the concept also involves processing on the user device. In Section \ref{evaluation}, we will prove that \emph{affinity}'s concept is realizable as conceived in Section \ref{concept}.

\subsection{Cold Start Problem}
The proposed concept requires an eco-system of users who have submitted their DF vectors to the backend, because we use tf$-$idf word relevance measurements. For example, if there were exactly two registered users, then the idf part of Equation (\ref{tfidf}) would either be zero for words both users submitted or $\log(2)$ else. This contradicts the initial purpose of finding relevant words since the word they have in common is evaluated to have zero relevance. Therefore, we will require a sufficiently large initial user base for the tf$-$idf features to work properly. For this purpose, we use the NUS SMS chat corpus\footnote{Dataset available at: https://doi.org/10.25540/WVM0-4RNX} \cite{Chen2013}, which contains about 40,000 English and 30,000 Mandarin Chinese SMS messages with mainly colloquial bits of conversations. We convert unique users' English messages into \emph{seed documents}. We then concatenate all seed documents into a big corpus and train an initial \emph{fasttext} reference model with default settings (in particular embedding dimension $D=100$), which is then kept secret. The \emph{fasttext} model then reflects linguistic properties of short colloquial messages. 

\subsection{Text Selection}
Instead of submitting personal texting histories, the backend fetches Facebook Messenger histories excluding messages not written by the user. We use a pre-trained \emph{fasttext} model\footnote{https://fasttext.cc/docs/en/language-identification.html\#content} for language classification. We drop conversations where the majority of words is non-English. The result is then appended to the user's reference document. 
\subsection{Latent Feature Pre-Calculation}
The user's BoW vector excluding stop words is updated. Then the global DF vector is updated. In order to avoid comparison on words containing typos or user-related artifacts, we exclude words from comparison if they occur in less than $p_{min} = 0.05$ reference documents which we found to perform well. For the remainder of the words, we update the tf$-$idf values in the BoW vector. We select the top $k=50$ tf$-$idf valued words as signature vectors and generate signature word vector representations via the initially trained \emph{fasttext} reference model. The signature word vectors together with their corresponding normalized BoW vector form the user's signature. Note that the size of a user signature is approximately 10 kB for the signature size $k=50$ on half-precision floats.

\subsection{Similarity Comparison}
When a user scans another user's id encoded as a QR code, both user ids are sent to the backend, their pre-calculated signatures are passed to the WMD. The result distance score is returned to both users and displayed on their smartphones. Note that the Word Mover's Distance (WMD) equipped with the cosine distance is bound between zero and two, where zero denotes perfect similarity and two perfect dissimilarity. Therefore, we renormalize as follows
\begin{equation}\label{eq:sim}
\text{sim($id_1$, $id_2$) = } 1- \frac{1}{2} \text{ WMD(($M_1$,$B_1$),($M_2$,$B_2$))},
\end{equation}
where WMD denotes the Word Mover's Distance equipped with the cosine distance, $M_i$ and $B_i$ denote the $i$-th id's signature word vector matrix and the normalized BoW vector respectively. Now similarity scores are bound between zero and one with one denoting perfect similarity and zero perfect dissimilarity. 

\section{Evaluation}\label{evaluation}

\subsection{Datasets}
We used the NUS SMS corpus in order to derive necessary seed users in order to be able to apply tf$-$idf relevance measures in a meaningful way. Furthermore, we used its concatenated English messages in order to train a \emph{fasttext} word embedding model, which is able to reflect linguistic properties of language used in short and colloquial messages. 

Tweets differ from SMS messages in for example their average words per message, the fact that SMS messages are likely to contain only a single word, or that tweets are publicly available and therefore often contain more structured and non-sensitive information \cite{Chen2013}. However, their short character limitations to 280 and 160 respectively create a common ground for shortened forms and abbreviations of words, or pictograms in both formats. Furthermore, since we use users' entire histories instead of single (potentially single-word) messages, we deem the impact of structural differences between tweet and SMS histories negligible in our scenario. Hence, we posit that tweet histories contain linguistic information similar to that of SMS histories or histories from any other (short) messaging service such as \emph{Facebook Messenger} or \emph{WhatsApp}. 

For evaluation, we use plain text tweets -- excluding metadata such as timestamps, number of retweets, or likes -- published until mid September 2018 by 60 US senators (30 Republicans, 30 Democrats) of the 115th US Congress\footnote{https://www.socialseer.com/resources/us-senator-twitter-accounts/}\footnote{Our dataset is available at: https://doi.org/10.7910/DVN/NMT4HP}. We drop stop words and web links.

\subsection{Performance}

\begin{figure}[htbp]
	\includegraphics[width=\linewidth]{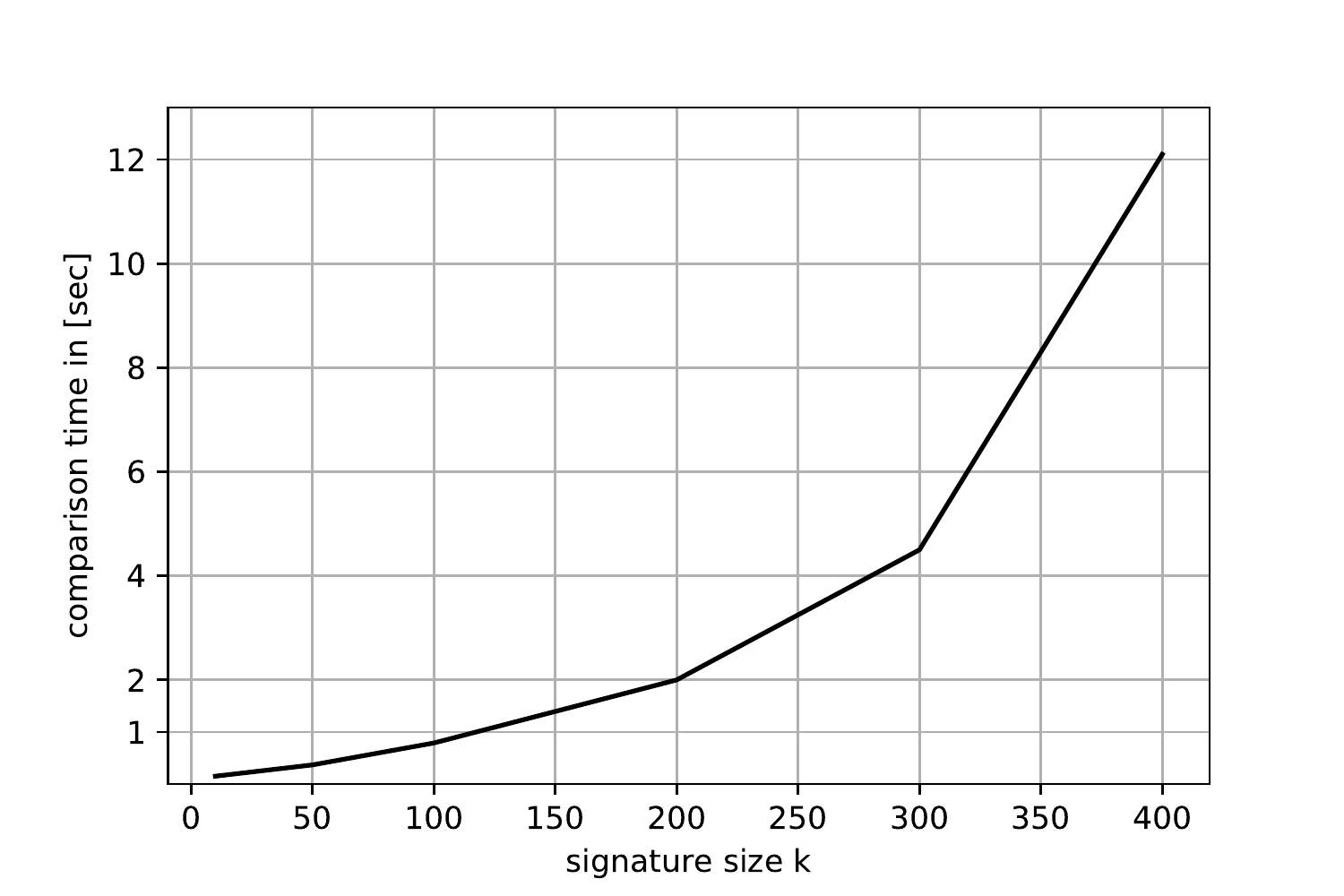}
	\caption{Average comparison times on a single CPU core (\emph{i7-3520M}) between Twitter reference accounts of 60 US senators subject to the signature size $k$.}
	\label{fig:performanceK}
\end{figure}

\emph{affinity} is designed to pre-calculate user signatures, that is a numerical condensate of linguistic information hidden in private texting data. We already know that the WMD scales super cubic in the signature size $k$. We benchmark calculation times of the \emph{affintiy} prototype on a single CPU core (\emph{i7-3520M}). The results are shown in Figure \ref{fig:performanceK}. We observe that the comparison time decreases to less than a second for $k\leq 100$, whereas for $k\geq 400$ comparison times exceed 10 seconds on average. In order to enable similarity comparison in a real-time scenario, comparisons should not take more than a second, yet without jeopardizing the quality of the similarity score. We will see that signature sizes of as small as $k=50$ yield good discriminative power. 

\subsection{Similarity Networks}
\begin{figure}[h]
	\includegraphics[width=\linewidth]{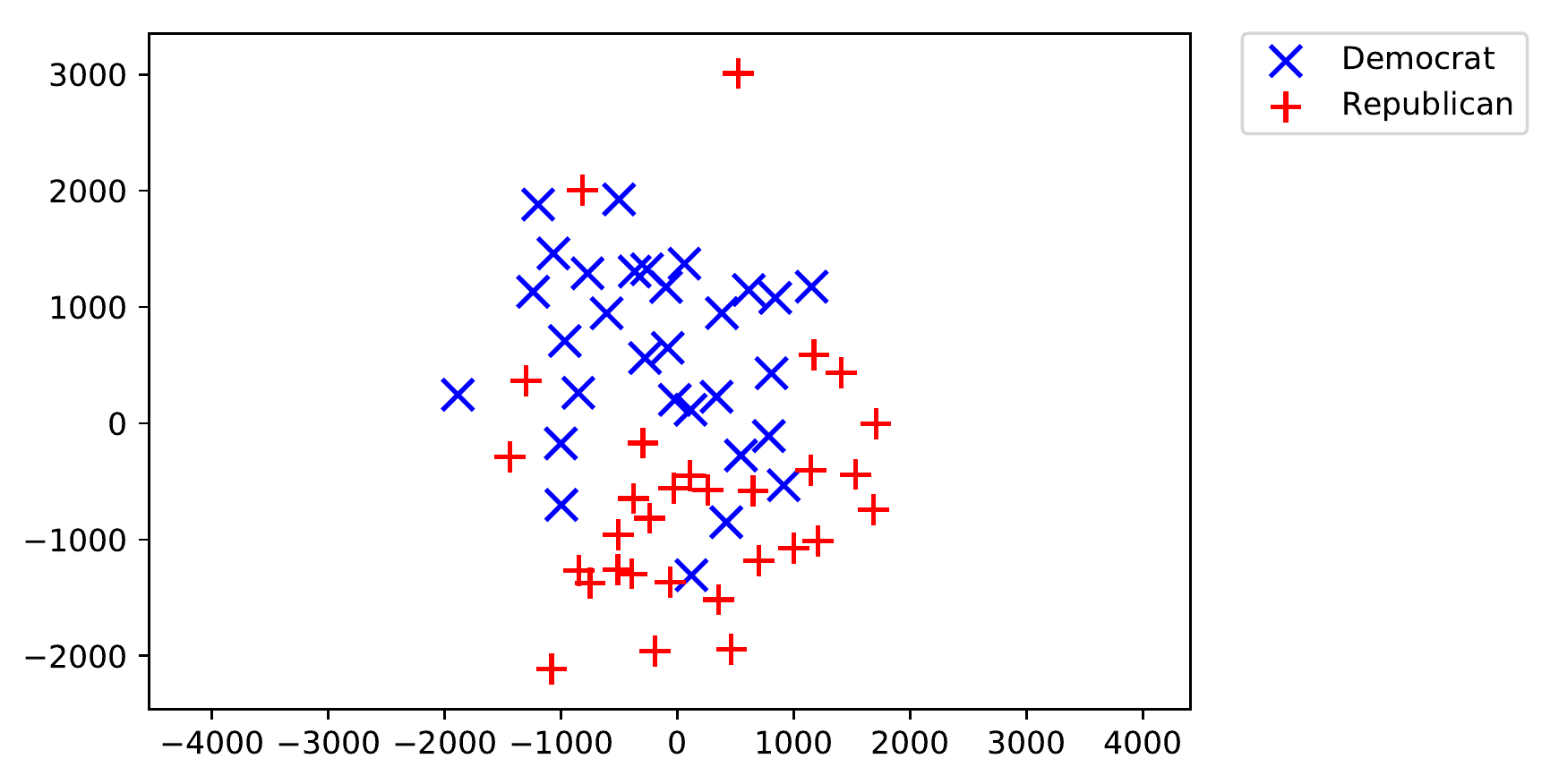}
	\caption{MDS visualization of pair-wise similarities based on $k=50$ signature words. Nodes represent 60 Twitter reference user accounts of US senators.}
	\label{fig:simNet}
\end{figure}

We define \emph{similarity networks} as a fixed set of fully connected users (nodes), where every edge is weighted by its corresponding pairwise distance. We call the corresponding weight matrix \emph{similarity matrix}. Multi-dimensional Scaling (MDS) \cite{Torgerson1952} allows to visualize the relative similarities between all nodes of the graph in two dimensions. MDS approximately preserves relative distances, in particular very similar users will be mapped to points close to each other. Since MDS requires a matrix of pairwise \emph{dissimilarities} instead of a similarity matrix, we transform similarities into dissimilarities via the mapping $x \mapsto 1-x$. An example MDS output for the signature size $k=50$ is depicted in Figure \ref{fig:simNet}.

Applying labels for Republican and Democratic party membership reveals that members of the same political party are mapped closer to each other. In order to measure \emph{affinity}'s quality of separating Democrats from Republicans, we propose a knowledge-based classifier following the principle of nearest neighbors. The classifier predicts the political party of a node by taking the majority vote over all its $n$ nearest neighbors. The results are shown in Table \ref{tab:simStability}. Note that a random classifier would predict party membership with an average accuracy of 50.0\% (30 Republicans : 30 Democrats).

\begin{table}[h]
	\centering
	\caption{\label{tab:simStability}Accuracy of predicting the political party correctly by majority vote on $n$ nearest neighbors of Twitter reference users of 60 US senators (30 Republicans, 30 Democrats) over several signature sizes $k$.}
	\label{tab:default}
		\begin{tabular}{@{}rrrr@{}}
			{} & \multicolumn{3}{c}{\textbf{\emph{k}}}  \\
			\cmidrule(l){2-4}
			\textbf{\emph{n}} &  10 & 50 & 100 \\
			\midrule
			1  & 0.733 & \textbf{0.850} &         0.817\\
			3  & 0.733 & \textbf{0.876} &         0.833\\
			5  & 0.750 &         0.850  & \textbf{0.867}\\
			7  & 0.783 & \textbf{0.850} & \textbf{0.850}\\
			9  & 0.783 & \textbf{0.833} &         0.816\\
			\midrule
			Avg. & 0.756 & \textbf{0.850} & 0.837\\
		\end{tabular}
		\\[6pt]
\end{table}

We see that the similarities derived from signature sizes as small as $k=50$ allow to predict the political party with an average accuracy of 85.0\%. \emph{affinity} is thus able to separate Democrat from Republican rhetorics based on their tweets without having seen any tweet history explicitly. In particular, the backend does not know in which regard users are similar since signatures are exchanged between users only. A user's signature data is incomprehensible for other users since it is ($a$) shared in latent form, and ($b$) the central word embedding model is not published. In that sense, \emph{affinity} is privacy-preserving.

\section{Conclusion}\label{conclusion}
In social networks, similar people tend to form links. On smartphone data, similarity is traditionally measured on sensor data, personal context data, or the social graph. We propose \emph{affinity}, a system for latent -- in particular privacy-preserving -- user similarity comparison based on texting data. Texting data stays on the smartphone and only a list of used words is disclosed in order to find individually relevant words. We select a small set of relevant words that are represented by latent word vectors and their corresponding word weights, which forms the necessary information for comparison via the Word Mover's Distance. Upon a local exchange of comparison data, similarity comparisons can be performed on the smartphone quickly and reliably without the backend knowing. \emph{affinity} does not assess similarity on interpretable text features yet on individual latent word embedding features. Comparison data in this latent format is incomprehensible to other users as long as the reference word embedding model is kept secret. Sharing comparison data is thus uncritical. In this sense, \emph{affinity} is privacy-preserving and allows to safely leverage the information potential hidden in private texting data. We evaluate our approach on a political party classification task and reach an average accuracy of 85.0\%.
\section{Future Work}\label{future_work}

We evaluated \emph{affinity} on a political party classification task. Evaluating \emph{affinity}'s performance on for instance gender or age classification tasks will substantiate \emph{affinity}'s generalizability and applicability.
We did not measure \emph{affinity}'s similarity comparison performance on smartphones. The evaluation of a smartphone implementation is due.

\bibliographystyle{IEEEtran}
\bibliography{similaritybib18}


\end{document}